\documentclass[11pt]{article}
\newcommand{\nc}{\newcommand}
\nc{\bib}{\bibitem}
\nc{\al}{\alpha}
\nc{\g}{\gamma}
\nc{\G}{\Gamma}
\nc{\D}{\Delta}
\nc{\eps}{\epsilon}
\nc{\la}{\lambda}
\nc{\La}{\Lambda}
\nc{\var}{\varphi}
\nc{\hn}{h^\vee}
\nc{\kn}{k^\vee}
\nc{\adg}{a^\dagger}
\nc{\bdg}{b^\dagger}
\nc{\ba}{\beta_\al}
\nc{\ga}{\g^{\al_1}}
\nc{\vpp}{{V_+}^+}
\nc{\cpp}{{C_+}^+}
\nc{\Vm}{V_{-\al^-}^{\al_1}}
\nc{\Vp}{V_{-\al^+}^{\al_1}}
\nc{\Vmb}{V_{-\beta^-}^{\al_1}}
\nc{\Gb}{\overline{G}}
\nc{\Gbc}{\overline{{\cal G}}}
\nc{\pa}{\partial}
\nc{\nn}{\nonumber \\ }
\nc{\hf}{\frac{1}{2}}         
\nc{\dz}{\frac{dz}{2\pi i}}
\nc{\fabc}{{f_{a,b}}^c}
\nc{\bin}[2]{\left (\begin{array}{c} {#1}\\ {#2} \end{array}\right )}
\nc{\ben}{\begin{equation}}
\nc{\een}{\end{equation}}
\nc{\bea}{\begin{eqnarray}}
\nc{\eea}{\end{eqnarray}}
\nc{\bra}[1]{\langle {#1}|}
\nc{\ket}[1]{|{#1}\rangle}
\newcommand{\Z}{\mbox{$Z\hspace{-2mm}Z$}}
\nc{\C}{\mbox{\hspace{1.24mm}\rule{0.2mm}{2.5mm}\hspace{-2.7mm} C}}
\nc{\Nat}{\mbox{\hspace{.04mm}\rule{0.2mm}{2.8mm}\hspace{-1.5mm} N}}
\newcommand{\R}{\mbox{\hspace{.04mm}\rule{0.2mm}{2.8mm}\hspace{-1.5mm} R}}
\nc{\spa}{\hspace{.1cm},\hspace{1 cm}}
\nc{\vs}{\vspace}
\nc{\NP}[1]{Nucl.\ Phys.\ {\bf #1}}
\nc{\PL}[1]{Phys.\ Lett.\ {\bf #1}}
\nc{\CMP}[1]{Commun.\ Math.\ Phys.\ {\bf #1}}
\nc{\PR}[1]{Phys.\ Rev.\ {\bf #1}}
\nc{\PRL}[1]{Phys.\ Rev.\ Lett.\ {\bf #1}}
\nc{\PTP}[1]{Prog.\ Theor.\ Phys.\ {\bf #1}}
\nc{\PTPS}[1]{Prog.\ Theor.\ Phys.\ Suppl.\ {\bf #1}}
\nc{\MPL}[1]{Mod.\ Phys.\ Lett.\ {\bf #1}}
\nc{\IJMP}[1]{Int.\ Jour.\ Mod.\ Phys.\ {\bf #1}}
\nc{\IM}[1]{Invent.\ Math.\ {\bf #1}}
\nc{\SJNP}[1]{Sov. J. Nucl. Phys.\ {\bf #1}}
\nc{\JHEP}[1]{J.\ High\ Energy Phys.\ {\bf #1}}

\def\tri#1#2#3#4#5#6#7#8#9{\matrix{\quad\cr #4\cr
	#3\quad#5\cr #2~\qquad #6\cr #1\quad #9\quad#8\quad#7\cr 
         \quad\cr}}

\def\vvdots{\mathinner{\mkern1mu\raise1pt\vbox{\kern7pt\hbox{.}}\mkern2mu
 \raise4pt\hbox{.}\mkern2mu\raise7pt\hbox{.}\mkern1mu}}

\begin{document}

\oddsidemargin 5mm

\begin{titlepage}
\setcounter{page}{0}

\vs{8mm}
\begin{center}
{\huge $su(N)$ tensor product multiplicities}\\[.2cm]
{\huge and}\\[.4cm]
{\huge virtual Berenstein-Zelevinsky triangles}

\vs{15mm}
{\large J. Rasmussen}\footnote{rasmussj@cs.uleth.ca; supported in part
by a PIMS Postdoctoral Fellowship and by NSERC} and 
{\large M.A. Walton}\footnote{walton@uleth.ca; supported in part by NSERC}
\\[.2cm]
{\em Physics Department, University of Lethbridge,
Lethbridge, Alberta, Canada T1K 3M4}

\end{center}

\vs{8mm}
\centerline{{\bf{Abstract}}}
\noindent
Information on $su(N)$ tensor product multiplicities is neatly encoded
in Berenstein-Zelevinsky triangles. Here we study a generalisation
of these triangles by allowing negative as well as non-negative 
integer entries. For a fixed triple product of weights, these
generalised Berenstein-Zelevinsky triangles span a lattice in which one may
move by adding integer linear combinations of so-called virtual triangles. 
Inequalities satisfied by
the coefficients of the virtual triangles describe a polytope. 
The tensor product multiplicities may be computed as the number
of integer points in this convex polytope. As our main result, we 
present an explicit formula for this discretised volume as a multiple sum.
As an application, we also address the problem of determining when a tensor
product multiplicity is non-vanishing. The solution is represented by a set
of inequalities in the Dynkin labels. We also allude to the question of when
a tensor product multiplicity is greater than a given non-negative
integer.

\end{titlepage}
\newpage
\renewcommand{\thefootnote}{\arabic{footnote}}
\setcounter{footnote}{0}

\section{Introduction}

The decomposition of tensor products of modules of simple Lie algebras 
has been studied for a long
time now. Many elegant results have been found for the multiplicities 
of the decompositions, the so-called tensor product multiplicities.  
The relatively recent Berenstein-Zelevinsky method of 
triangles \cite{BZ} is an example. 
Although it is a powerful, symmetric method, it is not explicit: 
triangles are constructed according to certain rules, and their number
is the required tensor product multiplicity. 
Here we show that a generalisation of these 
Berenstein-Zelevinsky (BZ) triangles allows us to work out a very
explicit expression for the multiplicities.
A tensor product multiplicity is expressed as a multiple sum, counting 
the number of integer points in a particular
convex polytope, to be defined below. BZ triangles and our results pertain
to the $A$-series; $A_r=su(r+1)$. We will sometimes write $su(N)$ with
$N=r+1$.

We are interested in describing decompositions of tensor products of
irreducible highest weight modules of simple Lie algebras. 
They are usually written 
\ben
 M_\la\otimes M_\mu=\bigoplus_\nu\ {T_{\la,\mu}}^\nu M_\nu\ \ ,
\label{MM}
\een
where $M_\la$ is the module of highest weight $\la$. ${T_{\la,\mu}}^\nu$ is
the tensor product multiplicity. We shall study the equivalent but more 
symmetric problem of determining the multiplicity of the singlet 
in the expansion of the triple product
\ben
 M_\la\otimes M_\mu\otimes M_\nu\supset T_{\la,\mu,\nu}M_0\ .
\label{MMM}
\een
If $\nu^+$ denotes the highest weight conjugate to 
$\nu$, we have $T_{\lambda,\mu,\nu} = {T_{\lambda,\mu}}^{\nu^+}$.  
We will use the shorthand notation $\la\otimes\mu\otimes\nu$ to represent the
left hand side of (\ref{MMM}).

An $su(3)$ BZ triangle, describing a particular coupling (to the singlet)
associated to the
triple product $\lambda\otimes\mu\otimes\nu$, is a triangular arrangement
of 9 non-negative integers:
\ben
 \matrix{\quad\cr m_{13}\cr
	n_{12}~~\quad l_{23}\cr
 m_{23}~\quad\qquad ~~m_{12}\cr
 n_{13}~\quad l_{12} \qquad n_{23} \quad~ l_{13} \cr \quad\cr}
\label{trithree}
\een
These integers are related to the Dynkin labels of the three integrable
highest weights by 
\ben
\begin{array}{llll}
 &m_{13}+n_{12}=\lambda_1\ ,\ \ &n_{13}+l_{12}=\mu_1\ ,\ \
 &l_{13}+m_{12}=\nu_1\ ,\nn
 &m_{23}+n_{13}=\lambda_2\ ,\ \ &n_{23}+l_{13}=\mu_2\ ,\ \
 &l_{23}+m_{13}=\nu_2\ .
\end{array}
\label{outthree}
\een 
We call these relations outer constraints.
The entries further satisfy the so-called hexagon conditions 
\ben
\begin{array}{l}
 n_{12}+m_{23}=n_{23}+m_{12}\ ,\nn 
 m_{12}+l_{23}=m_{23}+l_{12}\ ,\nn 
 l_{12}+n_{23}=l_{23}+n_{12}
\end{array}
\label{hexthree}
\een
of which only two are independent.
They say that the length of opposite sides of the
hexagon must be equal, if the length of a segment is defined 
to be the sum of the two integers associated to its
endpoints. An $su(3)$ BZ triangle is thus composed of one hexagon and
three corner points.

For $su(4)$ the BZ triangle is defined in a similar way, in terms of
18 non-negative integers:
\ben
 \matrix{m_{14}\cr
	n_{12}~~\quad l_{34}\cr
 m_{24}~\qquad\qquad ~~m_{13}\cr
 n_{13}\qquad l_{23}\qquad n_{23} \qquad l_{24}\cr
 m_{34}\qquad\qquad\quad m_{23}\qquad\qquad\quad m_{12} \cr
 n_{14}\qquad l_{12}\qquad n_{24}\ \ \quad l_{13}\quad~~~ n_{34}\qquad
 l_{14} \cr  }
\label{trifour}
\een
related to the Dynkin labels by
\ben
 \begin{array}{llll}
 &m_{14}+n_{12}=\lambda_1\ ,\ \ &n_{14}+l_{12}=\mu_1\ ,\ \
 &l_{14}+m_{12}=\nu_1\ ,\nn
 &m_{24}+n_{13}=\lambda_2\ ,\ \ &n_{24}+l_{13}=\mu_2\ ,\ \
 &l_{24}+m_{13}=\nu_2\ ,\nn
 &m_{34}+n_{14}=\lambda_3\ ,\ \ &n_{34}+l_{14}=\mu_3\ ,\ \
 &l_{34}+m_{14}=\nu_3\ .
\end{array}
\label{outfour}
\een
Furthermore, the $su(4)$ BZ triangle contains three hexagons:
\ben
 \begin{array}{llll}
 &n_{12}+m_{24} =m_{13}+n_{23}\ ,\ \  
 & n_{13}+l_{23} =l_{12}+n_{24}\ ,\ \
 & l_{24}+n_{23} =l_{13}+n_{34}\ ,\nn
 &n_{12}+l_{34}=l_{23}+n_{23}\ ,\ \
 & n_{13}+m_{34} =n_{24}+m_{23}\ ,\ \
 & n_{23}+m_{23} =m_{12}+n_{34}\ ,\nn
 &m_{24}+l_{23} =l_{34}+m_{13}\ ,\ \
 & m_{34}+l_{12} = l_{23}+m_{23}\ ,\ \
 &l_{13}+m_{23} =l_{24}+m_{12}\ . 
\end{array}
\label{hexfour}
\een
It is a general feature for any $N$ that only two out of the three hexagon 
identities associated to a particular hexagon are independent.

The $su(N)$ generalisation is obvious; the triangle is built out of
$(N-1)(N-2)/2$ hexagons and three corner points. Simple examples of 
lower rank BZ triangles and their applications may be found
in Ref. \cite{FMS}.

\section{Generalised and virtual Berenstein-Zelevinsky triangles}

The generalisation of the BZ triangles we shall consider is obtained by
weakening the constraint that all entries are {\it non-negative} integers to
{\it arbitrary} 
integers, negative as well as non-negative. The hexagon identities
and the outer constraints are still enforced. A triangle will be called a
{\it true} BZ triangle if all its entries are non-negative.

We consider a generalised BZ triangle associated to $su(r+1)$.
Denoting the number of entries $E_r$ and the number of hexagons $H_r$,
we have 
\ben
 E_r=\frac{3}{2}r(r+1)\ ,\ \ \ \ \ \ H_r=\frac{1}{2}r(r-1)\ .
\label{EH}
\een
For a given triple product $\lambda\otimes\mu\otimes\nu$, the set of associated
triangles spans an $H_r$-dimensional lattice. 
Each hexagon corresponds to two independent constraints on the triangle 
entries
while there are $3r$ outer constraints. This leaves 
\ben
 E_r-(2H_r+3r)=H_r
\label{EHrel}
\een
parameters labelling the possible triangles.
Among these, only a finite number are true BZ triangles. This number is
precisely the tensor product multiplicity of the triple coupling.
For example, when the singlet does not occur in the decomposition of
the triple product, there are no true BZ triangles
in the lattice. 

A special class of generalised BZ triangles is 
associated to the triple product $0\otimes0\otimes0$. We say they have 
weight $(\lambda,\mu,\nu) = (0,0,0)$.  
According to the general argument above, $H_r=\frac{1}{2}r(r-1)$ 
such triangles are linearly independent. 
We shall call them {\it virtual} triangles, and denote them using ${\cal V}$.
It is natural to exclude the triangle with all entries equal to zero
from the set of virtual triangles. It is the unique true BZ triangle
in the lattice associated to the triple product $0\otimes0\otimes0$.
In the cases of $su(3)$ and $su(4)$, virtual triangles have appeared in 
Ref. \cite{BMW} and Ref. \cite{BKMW}, respectively.

A convenient basis for the virtual triangles is given by associating
a simple distribution of plus and minus ones 
(written $1$ and $\bar{1}\equiv-1$,
respectively) to a given hexagon. All other entries are zero. The
distribution is
\ben
 \matrix{\matrix{1\cr 1\quad\bar1~~\quad\bar1\quad 1\cr
  \bar1~\quad\quad ~~\bar1\cr
  1~\quad\bar1~\quad\bar1~\quad 1\cr
 1\cr}}
\label{virt}
\een
Thus, a basis virtual triangle will always
have 6 entries equalling $-1$, and between 3 and 6 
entries equalling $+1$. The number of $+1$ entries depends on where
the associated hexagon is situated in the generalised BZ triangle and
on the rank of $su(r+1)$ (3 entries equal $+1$ for $su(3)$ only). That these 
virtual triangles are indeed linearly independent is obvious.

There are no virtual triangles in the case of $su(2)$.
In the case of $su(3)$ there is one basis virtual triangle
\ben
 {\cal V}=\tri{1}{\bar1}{\bar1}{1}{\bar1}{\bar1}{1}{\bar1}{\bar1}
\label{Vthree}
\een
while in the case of $su(4)$ the three basis virtual triangles 
${\cal V}_1$, ${\cal V}_2$ and ${\cal V}_3$ are
\bea
 {\cal V}_1=\matrix{1\cr
	\bar1~~\quad\bar1\cr
 \bar1~\quad\quad ~~\bar1\cr
 1~\quad\bar1\quad\bar1\quad ~1\cr
 0\qquad\quad 1\qquad\quad 0 \cr
 0~\quad 0~\quad 0~\quad 0~\quad 0~\quad 0 \cr}
\hspace{1cm} {\cal V}_2=\matrix{0\cr
	0~~\quad 0\cr
 1~\quad\quad ~~0\cr
 \bar1~\quad \bar1\quad 1 \quad ~0\cr
 \bar1\qquad\quad \bar1\qquad\quad 0 \cr
 1~\quad\bar1~\quad\bar1~\quad 1~\quad 0~\quad 0 \cr}\nn
 {\cal V}_3=\matrix{0\cr
	0~~\quad 0\cr
 0~\quad\quad ~~1\cr
 0~\quad 1\quad\bar1 \quad ~\bar1\cr
 0\qquad\quad\bar1\qquad\quad\bar1 \cr
 0~\quad 0~\quad 1~\quad\bar1~\quad\bar1~\quad 1\cr}\hspace{3cm}\mbox{}
\label{Vfour}
\eea
The generalisation to higher rank $su(r+1)$ is straightforward.
In Section 3 we shall use another choice of indices on ${\cal V}$.

We are now in a position to generate all generalised BZ triangles 
associated to a given triple coupling. Once a single generalised 
triangle has been found, the lattice
of triangles associated to the triple coupling is spanned
by adding integer linear combinations of the virtual triangles. 
The choice of initial triangle
is not important. We shall denote the integer coefficients 
{\it linear coefficients}.
We emphasise that the euclidean spaces spanned by the lattices are all of 
dimension $H_r$, i.e., the dimension is independent of the triple
coupling and depends only on the rank of $su(r+1)$.

Negative entries in (generalised) BZ triangles appear also in Ref. 
\cite{BCM}.  
That work is mainly devoted to the construction of tensor-product generating
functions. A new method is proposed based on elementary solutions
to certain sets of linear equations related to the BZ triangles, and for
$su(3)$ one of these solutions corresponds to a triangle with negative
entries. The appearance of negative entries is expected to be a general
feature for higher $su(N)$ as well. 
The elementary solutions are closely related to the
so-called elementary couplings. 

\section{Polytopes, multiple sums and tensor product multiplicities}

We shall now focus on the $H_r$-dimensional linear coefficient space and
seek an algebraic description of the tensor product multiplicities.
The latter are computed by counting true BZ triangles. Demanding that 
the entries of a {\it true} BZ triangle be non-negative, we 
obtain $E_r$ inequalities 
the linear coefficients must satisfy. The inequalities 
depend on the choice of initial triangle, and they correspond to
a polyhedral combinatorial 
expression for the multiplicities in linear coefficient space.
The structure of the basis virtual triangles (cf. (\ref{virt}), 
(\ref{Vthree}) and (\ref{Vfour})) ensures that all
linear coefficients have upper as well as lower bounds. The polyhedron is
therefore bounded and such a polyhedron is called a polytope. It is easily
seen to be convex. 

In order to specify the polytope, we must find an initial triangle of 
weight $(\lambda,\mu,\nu)$. It is convenient to break the symmetry 
among the 3 weights, and first look at the unique true triangle of 
weight $(\lambda,\mu,\lambda^++\mu^+)$. For $su(3)$, this triangle is 
\ben
 \matrix{\quad\cr \lambda_1\cr
	0~~\quad \mu_1\cr
 \lambda_2~\quad\qquad ~~\lambda_2\cr
 0~\quad \mu_1 \qquad 0 \quad~ \mu_2 \cr \quad\cr}
\label{hitri}
\een
and the generalisation to $su(r+1)$ is clear. Every highest weight $\nu$ in 
a coupling $\lambda\otimes\mu\otimes\nu$ satisfies 
\ben
 \nu\ =\ \lambda^++\mu^+ -\sum_{i=1}^rn_i\alpha_i\ ,
\label{Nal}
\een
with $n_i\in\Z_\ge$, 
where $\alpha_i$ is the $i$-th simple root. The coefficients $n_i$ are 
conveniently expressed using dual Dynkin labels. A weight $\lambda$ can be 
written 
\ben
 \la\ =\ \sum_{i=1}^r\la_i\Lambda^i\ =\ \sum_{i=1}^r\la^i\al_i^\vee\ , 
\label{Dynkin}
\een
where $\{\Lambda^i\}$ and $\{\al^\vee_i\}$ are the sets of fundamental
weights and simple co-roots, respectively. 
The $\la^i$ are the dual Dynkin labels, while the ordinary
Dynkin labels are the $\la_i$. For simply-laced algebras, 
like $su(N)$, $\alpha_i$ is identical to the co-root $\alpha_i^\vee$ (with  
standard normalisation $\alpha^2 = 2$, for $\alpha$ a long root). 
Taking the scalar product of (\ref{Nal}) with $\Lambda^i$  
therefore gives 
\ben
 n_i  \ =\ (\lambda^+)^i +(\mu^+)^i - \nu^i\ .
\label{Nis}
\een
Generalised triangles of weight $(0,0,\alpha_i)$ are also easily 
constructed. An $su(3)$ example is
\ben
 \matrix{\quad\cr 1\cr
	\bar 1~~\quad 1\cr
 0~\quad\qquad ~~\bar1\cr
 0~\quad 0 \qquad 0 \quad~ 0 \cr \quad\cr}
\label{alii}
\een
of weight $(0,0,\alpha_2)$. So, one can find a generalised triangle 
of weight $(\lambda,\mu,\nu)$ by subtracting non-negative integer 
multiples of triangles of weight $(0,0,\al_i)$, such as (\ref{alii}), from 
a triangle like (\ref{hitri}), of weight $(\la,\mu,\la^++\mu^+)$. 

The result for $su(r+1)$ is the following generalised BZ 
triangle associated to the triple product $\la\otimes\mu\otimes\nu$: 
\ben
 \matrix{ N'_{r} \cr
      n_r \qquad N_{r} \cr
 \ \ \la_{2} \qquad\qquad\ \ \ \ N'_{r-1}  \cr
 \ \ 0 \ \qquad \mu_{1} \qquad n_{r-1} \quad N_{r-1} \cr
 \ \la_{3} \quad\qquad\qquad\ \ \la_{3} \quad\qquad\qquad N'_{r-2} \cr
 \ \ 0\ \qquad\ \mu_{1}\qquad 0\qquad\ \ \ \mu_2\quad\ \ n_{r-2}
   \ \ \ N_{r-2} \cr
 \ \vvdots \quad\qquad\qquad\qquad \vdots\qquad\quad \ \
  \vdots\qquad\qquad\qquad\quad \ddots\ \ \  \cr
 \la_{r-2}  \qquad\qquad\qquad\qquad\qquad\quad
   \qquad\qquad\qquad\qquad\qquad  N'_{3} \ \      \cr
 0\ \qquad\ \mu_{1}  \qquad\qquad\qquad\qquad\qquad\qquad\qquad\qquad
   \qquad\quad \     n_{3}\qquad N_{3} \ \   \cr
 \la_{r-1}\qquad\qquad\ \la_{r-1} \qquad\qquad\qquad\ \dots\qquad 
  \qquad\qquad\quad 
    \la_{r-1}\ \qquad\qquad\ N'_{2}\quad  \cr
 0\qquad\ \mu_{1}\qquad 0\qquad\ \mu_{2}
   \qquad\qquad\qquad\qquad\qquad \quad\qquad\
  0\qquad \mu_{r-2}\qquad n_{2}\ \ \ \quad N_{2} \    \cr  
 \ \la_{r}\qquad\qquad\quad\ \la_{r}\qquad\qquad\quad \la_{r}
    \qquad\qquad\qquad\qquad\quad \ \ 
  \la_{r}\qquad\qquad\quad\ \la_{r}\qquad\quad\quad\ \ \ N'_{1}\quad \cr  
 0\qquad\ \mu_{1}\qquad 0\quad~\quad \mu_2\quad~~~ 0\qquad\ \ \mu_{3} 
   \qquad\quad\dots\qquad\quad
 0\qquad\mu_{r-2}\ \ \ \ 0\quad\ \ \
  \mu_{r-1}\quad\ \ n_{1}\quad\ \ N_{1}\ \ \cr\cr}
\label{initial}
\een
The entries $n_i$, $N_i$ and $N'_i$ are defined by
\bea
 n_i&=&\la^{r-i+1}+\mu^{r-i+1}-\nu^i\ ,\nn
 N_i&=&(1-\delta_{i1})n_{i-1}-n_i+\mu_{r-i+1}\nn
  &=&-\la^{r-i+1}+(1-\delta_{i1})
  \la^{r-i+2}-(1-\delta_{ir})\mu^{r-i}
  +\mu^{r-i+1}-(1-\delta_{i1})\nu^{i-1}+\nu^i\ ,\nn
 N'_i&=&\nu_i-N_i\nn
  &=&\la^{r-i+1}-(1-\delta_{i1})\la^{r-i+2}+(1-\delta_{ir})\mu^{r-i}
  -\mu^{r-i+1}+\nu^i-(1-\delta_{ir})\nu^{i+1}\ .
\label{nN}
\eea

In order to be able to describe the polytope explicitly, we need to label 
the virtual triangles. Our choice is to write them as ${\cal V}_{i,j}$
or ${\cal V}_l$ depending on where the associated hexagons are situated.
The corresponding linear coefficients are denoted $d_{i,j}$ and $\eta_l$:
\ben
\matrix{\star\cr
	\star~~\qquad \star\cr
 \star~\qquad \eta_{r-1}\quad ~~\star\cr
 \star\qquad\ \star\qquad\ \ \star\ \qquad \star\cr
 \star\quad\ d_{r-2,1}\quad \star\qquad\eta_{r-2}\quad\ \star \cr
 \star\quad\ \ \star\qquad \ \star\quad~~\quad \star\qquad\ \star\quad\ \ 
  \star \cr
 \vvdots\ 
  \qquad\qquad\qquad\vdots\ \ \ \qquad\vdots\qquad\qquad\qquad\ddots\cr
 \star \quad\qquad\qquad\qquad\qquad\quad
   \qquad\qquad\qquad\qquad\qquad \star\cr
	\star~~\qquad \star \ \quad\qquad\qquad\qquad\qquad\quad
   \qquad\qquad\qquad \star~~\qquad \star\cr
 \star~\qquad d_{2,1}\quad ~~\star
   \qquad\qquad\qquad \dots \qquad\qquad\qquad
 \star~\qquad \eta_{2}\ \quad ~~\star\cr
 \star\quad\ \ \star~~\qquad \star \ \qquad \star 
    \qquad\qquad\qquad\qquad\qquad
 \star\qquad\ \star~~\qquad \star\ \ \quad \star\cr
 \star\qquad d_{1,1} \ \quad \star\qquad d_{1,2}\ \ \quad \star 
     \qquad\quad\qquad\qquad\
 \star\quad\ \ d_{1,r-2}\quad \star\qquad\eta_{1}\ \ \ \quad \star \cr
 \star\qquad \star\ \ \quad \star~~\qquad \star\qquad \ \star\qquad~ 
  \star  
     \quad\ \dots \quad\
 \star\qquad~ \star\qquad \star~~\qquad \star\qquad \star\qquad 
  \star \cr\cr}
\label{hexnot}
\een
Here a $\star$ indicates an unspecified 
entry while $d_{i,j}$ and $\eta_l$ are the linear
coefficients of the virtual triangles. They are depicted at 
the centres of the hexagons associated to the corresponding 
virtual triangles. We have chosen two 
different notations for the virtual triangles (and their associated
linear coefficients) to reflect the positions 
of the corresponding hexagons in 
the asymmetric initial triangle (\ref{initial}).

Now, denoting the initial triangle ${\cal T}_0$, any triangle in 
the lattice of general triangles of weight $(\la,\mu,\nu)$ 
may be written as 
\ben
 {\cal T}={\cal T}_0+\sum_{l=1}^{r-1}\eta_l{\cal V}_l+
  \sum_{i,j\geq1}^{i+j=r-1}d_{i,j}{\cal V}_{i,j}\ .
\label{T}
\een
The associated polytope of interest is in the $H_r$-dimensional
space spanned by $d_{i,j}$ and $\eta_l$. It is bounded by 
the inequalities
requiring that all entries in ${\cal T}$ are non-negative
(whereby ${\cal T}$ is ensured to be a pure BZ triangle). 
Hence, the position of the
polytope depends on the initial triangle ${\cal T}_0$. Nevertheless,
the {\it volume} of the polytope, the number of integer points bounded by the
polytope, is {\it independent} of ${\cal T}_0$.
By construction, this number is the tensor product multiplicity
$T_{\la,\mu,\nu}$ of the triple coupling $\la\otimes\mu\otimes\nu$ to
the singlet. 

Using the explicit choice of initial triangle (\ref{initial}) 
and the basis of virtual
triangles discussed above, it is simple to write down the
inequalities defining the polytope. To illustrate, we list
the three inequalities given by the three entries located furthest to the 
right on the bottom line of ${\cal T}$:
\ben
 \mu_{r-1}+d_{1,r-2}-\eta_1\geq0\ ,\ \ \ \ 
 n_1-\eta_1\geq0\ ,\ \ \ \ 
 N_1+\eta_1\geq0\ .
\label{ex}
\een

A similar polyhedral combinatorial expression is discussed in Ref.
\cite{GZ}. 
Convex polytopes constructed there lie in the space of Gelfand-Tsetlin 
patterns (see e.g. \cite{GZ,FMS}), 
while ours lie in spaces associated to BZ triangles.
Hence, for $su(N)$ their polytopes
are embedded in the euclidean vector space $\R^{N(N+1)/2}$, while
ours may be embedded in the smaller space $\R^{(N-2)(N-1)/2}$.
A more universal method of constructing polyhedral combinatorial expressions
for tensor product multiplicities, generalising that of Ref. \cite{GZ} and
making sense for any simple Lie algebra, 
may be found in Refs. \cite{BZmath}. 

\subsection{Explicit multiple sum formula}

As already stated, our polyhedral expression differs from the one 
discussed in Refs. \cite{GZ,BZmath}. 
Its structure allows us to extract an explicit 
multiple sum formula counting the integer points bounded by the polytope. 
The multiple sum is over the linear coefficients, so 
different orders of summation give  a total
of $H_r!$ possible representations of the polytope volume.
For practical purposes, however, there are considerably fewer appropriate
summation orders. 
Let us illustrate our procedure for choosing an appropriate order
of summation by considering the following simple planar example.

Let a planar polytope be defined by the set of inequalities
\ben
\begin{array}{lll}
 &1\leq x\leq4\ ,\hspace{1cm}&8\leq x+y\leq14\ ,\nn
 &6\leq y\ ,\ \  &4\leq y-x\leq8\ .
\label{planar}
\end{array}
\een
The volume or area ${\cal A}$ of the polytope
(the number of integer points bounded by the inequalities)
can be written in two ways:
\ben
 {\cal A}\ =\ 16\ =\ \sum_{x=1}^4\ 
 \sum_{y={\rm max}\{6,x+4,8-x\}}^{{\rm min}\{x+8,
  14-x\}}1\ =\ \sum_{y=6}^{11}\
 \sum_{x={\rm max}\{1,y-8,8-y\}}^{{\rm min}\{4,y-4,14-y\}}1\ .
\label{sumA}
\een
The second expression is slightly more difficult to write, since the
upper limit 11 on $y$ must be calculated from the intersection of 
faces (lines). Here the bounding lines $x+y=14$ and $y-x=8$
intersect at the point $(x,y)=(3,11)$. This is a complication to 
avoid when writing our formula, since it will involve many sums. 
In the first expression the explicitly written lower limit
$y=6$ is redundant since the remaining two intersect at the point $(2,6)$.
However, including redundant limits does not change the result, so we
may choose to keep the limit $y=6$.

It is clear that an important difference between the 
two orders of summation is that we have $1\le x\le 4$, but only 
$6\leq y$. For example, if (\ref{planar}) is supplemented with 
$y\le 15$, then the upper limit 11 can be replaced simply by 15, and 
the formula is still valid. While it is true that 
the new formula contains single sums with lower limits greater than 
upper limits, this is a relatively small inconvenience. Those sums 
simply contribute 0. 

So, when choosing an appropriate order of summation (over the summation 
variables $\eta_l$ and $d_{i,j}$), it is crucial that for any summation
variable, all subsequent summation variables have upper
as well as lower bounds parallel to (or independent of) the one under
consideration.

This is a non-trivial consideration: the procedure does not
apply to all polytopes. A trivial example is provided by (\ref{planar}) with 
the inequality $x\leq4$ removed.

Fortunately, the simplifying procedure applies to our polytope. A simple 
inspection reveals that not all summation orders are appropriate, however. 
Nevertheless, in the general case and in accordance with our procedure,
we may express the volume of the polytope as
\ben
 T_{\la,\mu,\nu}=\left(\sum_{d_{1,1}}\right)\left(\sum_{d_{2,1}}\sum_{d_{1,2}}
   \right)...\left(\sum_{d_{r-2,1}}...\sum_{d_{1,r-2}}\right)
    \left(\sum_{\eta_{r-1}}...\sum_{\eta_1}\right)1
\label{vol}
\een
where the summation variables are bounded according to
\bea
 &&{\rm max}\{-N_1,d_{1,r-2},-N_2'+\eta_2,-\mu_{r-2}
  +d_{1,r-2}-d_{2,r-3}+\eta_2\}\nn
 &&\quad\quad
  \leq\eta_1\leq{\rm min}\{n_1,\mu_{r-1}+d_{1,r-2},\la_r-d_{1,r-2}+\eta_2,
  n_2+d_{1,r-2}-\eta_2,N_1',N_2+\eta_2\}\ ,\nn
 &&{\rm max}\{d_{l-1,r-l}-d_{l-1,r-l-1}+d_{l,r-l-1},-N_{l+1}'
   +\eta_{l+1},\nn
 &&\quad\quad\quad-\mu_{r-l-1}+\eta_{l+1}
   -(1-\delta_{l,r-2})d_{l+1,r-l-2}+d_{l,r-l-1}\}\nn
 &&\quad\quad
  \leq\eta_{l}\leq{\rm min}\{\la_{r-l+1}-d_{l,r-l-1}
   +d_{l-1,r-l}+\eta_{l+1},\nn
 &&\hspace{3.2cm}n_{l+1}+d_{l,r-l-1}-\eta_{l+1},N_{l+1}+\eta_{l+1}\}\ ,\ 
  \ {\rm for}\ 2\leq l\leq r-2\ ,\nn
 &&{\rm max}\{d_{r-2,1},-N_r'\}
 \leq\eta_{r-1}\leq{\rm min}\{\la_2+d_{r-2,1},n_r,N_r\}\ ,\nn
 &&{\rm max}\{d_{1,j-1},-\mu_{j-1}+d_{1,j-1}+d_{2,j-1}-(1-\delta_{j,2})
  d_{2,j-2}\}\nn
 &&\quad\quad
   \leq d_{1,j}\leq{\rm min}\{\mu_{j}+d_{1,j-1},
  \la_r-d_{1,j-1}+d_{2,j-1}\}\ ,\ \ {\rm for}\ 2\leq j\leq r-2\ ,\nn
 &&{\rm max}\{d_{i,j-1}+d_{i-1,j}-d_{i-1,j-1},-\mu_{j-1}
  +d_{i,j-1}+d_{i+1,j-1}-(1-\delta_{j,2})d_{i+1,j-2}\}\nn
 &&\quad\quad\leq d_{i,j}\leq\la_{r-i+1}-d_{i,j-1}+d_{i+1,j-1}+d_{i-1,j}
  \ ,\ \ {\rm for}\ 2\leq i,j,i+j\leq r-1\ ,\nn
 &&d_{i-1,1}\leq d_{i,1}\leq
  \la_{r-i+1}+d_{i-1,1}\ ,\ \ {\rm for}\ 2\leq i\leq r-2\ ,\nn
 &&0\leq d_{1,1}\leq{\rm min}\{\mu_1,\la_r\}\ .
\label{bounds}
\eea
{}From (\ref{Nal}) and (\ref{Nis}) it follows that the weights are subject to
the condition
\ben
 \la^i+\mu^i+\nu^i\ \in\ \Z_\geq\ ,\ \ \ \ i=1,...,r\ ,
\label{inte}
\een
ensuring the integer nature of the entries, and thus also of the
summation limits (\ref{bounds}).
The multiple sum formula (\ref{vol}) is our main new result. We now 
demonstrate its usefulness by considering an application and working out
a few examples.

\section{An application}

It is of interest to know whether or not a coupling  of a certain weight 
$(\la,\mu,\nu)$ exists, without
having to work out the tensor product multiplicity. 
Based on our multiple sum formula
(\ref{vol}) and (\ref{bounds}) one may derive a set of inequalities in the
dual and ordinary Dynkin labels of the three weights, determining
when the associated tensor product multiplicity is non-vanishing.
To illustrate the method,  we discuss the inequalities
for $su(3)$ and $su(4)$ and outline their derivation. 
In principle, it is possible to repeat the procedure for higher
rank, but even for $su(4)$ the derivation is very cumbersome.
We believe that similar results exist for all simple Lie algebras, 
and hope to report more general results later.

The dimension of the linear coefficient space 
for $su(3)$ is one, so the tensor product multiplicity may be
represented by a single sum:
\ben
 T_{\la,\mu,\nu}=\sum_{\eta={\rm max}\{0,\la^2+\mu^1-\mu^2-\nu^1,
   -\la^1+\la^2+\mu^1-\nu^2\}}^{{\rm min}\{\mu_1,\la_2,\la^2+\mu^2-\nu^1,
  \la^1+\mu^1-\nu^2,-\la^1+\la^2+\mu^1-\nu^1+\nu^2,\la^2+\mu^1-\mu^2
  +\nu^1-\nu^2\}}1\ .
\label{Tthree}
\een
The weights are subject to the integer constraint (\ref{inte}).
Note that the summation limits are not symmetric in the weights.
This is simply because we have chosen an asymmetric initial triangle.
The summation (\ref{Tthree}) 
is non-vanishing if and only if the upper limit is greater than or
equal to the lower limit. This condition yields $6\cdot3=18$ inequalities:
\bea
 0&\leq&\la_i,\mu_i,\nu_i\ , \ \ \ \ \ {\rm for} \ \ i=1,2\ ,\nn
  {\rm max}\{\la^1-\la^2+\mu^1-\mu^2,-\la^1+\mu^2,\la^2-\mu^1\}&\leq
 &\nu^1\leq\la^2+\mu^2\ ,\nn
 {\rm max}\{-\la^1+\la^2-\mu^1+\mu^2,\la^1-\mu^2,-\la^2+\mu^1\}&\leq
 &\nu^2\leq\la^1+\mu^1\ ,\nn
 {\rm max}\{-\la^2-\mu^1+\mu^2,-\la^1+\la^2-\mu^2\}&\leq
 &\nu^1-\nu^2\nn
 &\leq&{\rm min}\{\la^1-\mu^1+\mu^2,-\la^1+\la^2+\mu^1\}\ .
\label{eqthree}
\eea

Expressing the inequalities in terms of ordinary Dynkin labels, one should
bear in mind that the summation variable increases in steps of one while
the quadratic-form matrix involves a factor of $1/3$. A similar
factor $1/N$ is present for higher rank $su(N)$.

In the case of $su(4)$ the tensor product multiplicity may be written 
as a triple sum:
\bea
 T_{\la,\mu,\nu}&=&\sum_{d=0}^{{\rm min}\{\mu_1,\la_3\}}\ \sum_{\eta_2=
  {\rm max}\{d,-\la^1+\la^2+\mu^1-\nu^3\}}^{{\rm min}\{\la_2+d,\la^1+\mu^1
  -\nu^3,-\la^1+\la^2+\mu^1-\nu^2+\nu^3\}}\nn
 &\times&\sum_{\eta_1={\rm max}\{\la^3+\mu^2-\mu^3-\nu^1,d,-\la^2+\la^3-\mu^1
   +\mu^2-\nu^2+\nu^3+\eta_2,-\mu_1+d+\eta_2\}}^{{\rm min}\{\la^3+\mu^3-\nu^1,
  \mu_2+d,\la_3-d+\eta_2,\la^2+\mu^2-\nu^2+d-\eta_2,\la^3+\mu^2-\mu^3
  +\nu^1-\nu^2,-\la^2+\la^3-\mu^1+\mu^2-\nu^1+\nu^2+\eta_2\}}1\ .\nn
 &{}& 
\label{Tfour}
\eea
The weights are subject to the integer constraint (\ref{inte}).
For a multiple sum like this, the inequalities are obtained by first
considering the interior summation over $\eta_1$, 
leading to $6\cdot4=24$ inequalities
which may depend on the remaining two summation variables: 4 of them
depend only on the weights, 6 depend on $d$ but not $\eta_2$, while
14 depend on $\eta_2$. Treating the latter in the same way as the
upper and lower bounds
on the $\eta_2$-summation, we obtain a total of $13\cdot6=78$ inequalities
from the $\eta_2$-consideration. Repeating the procedure for $d$ leads
to a total of $54\cdot10=540$ 
inequalities in addition to the ones already derived
from the $\eta_1$- and $\eta_2$-considerations. This huge set of inequalities
may be reduced considerably and we find the following constraints
on the Dynkin labels (expressed in terms of dual as well as ordinary ones)
\bea
 0\leq\la_i,\mu_i,\nu_i\ , \ \ \ \ \ {\rm for} \ \ i=1,2,3&,&\nn
 {\rm max}\{\la^3-\mu^1,\la^3-\la_3-\mu^1+\mu_1,-\la^1+\mu^3,-\la^1+\la_1
  +\mu^3-\mu_3\}&\leq&\nu^1\leq\la^3+\mu^3\ ,\nn
 {\rm max}\{|\la^2-\mu^2|,|\la^2-\la_2-\mu^2+\mu_2|,|\la^1-\la^3
  +\mu^1-\mu^3|\}&\leq&\nu^2\leq\la^2+\mu^2\ ,\nn
 {\rm max}\{\la^1-\mu^3,\la^1-\la_1-\mu^3+\mu_3,-\la^3+\mu^1,-\la^3+\la_3
  +\mu^1-\mu_1\}&\leq&\nu^3\leq\la^1+\mu^1\ ,\nn
 {\rm max}\{\la^2-\la^3-\mu^1,-\la^1+\mu^2-\mu^3,\la^1-\la^2+\mu^1-\mu^2\}
  &\leq&\nu^1-\nu_1\nn
 \leq{\rm min}\{\la^2-\la^3+\mu^3,\la^3+\mu^2-\mu^3\}&,&\nn
 {\rm max}\{-\la^1+\la^3
  +\mu_2-\mu^2,\la_2-\la^2-\mu^1+\mu^3,\la^1-\la^3
  +\mu_2-\mu^2,&&\nn
 \quad\la_2-\la^2+\mu^1-\mu^3,-\la^2+\mu^2-\mu_2,\la^2-\la_2-\mu^2\}&\leq& 
  \nu^2-\nu_2\nn
  \leq{\rm min}\{\la^2-\la_2+\mu^2,\la^2+\mu^2-\mu_2\}&,&\nn
 {\rm max}\{\la^2-\la^1-\mu^3,-\la^3+\mu^2-\mu^1,\la^3-\la^2+\mu^3-\mu^2\}
  &\leq&\nu^3-\nu_3\nn
 \leq{\rm min}\{\la^2-\la^1+\mu^1,\la^1+\mu^2-\mu^1\}&,&\nn
 {\rm max}\{-\la^1+\la^3-\mu^2,-\la^2-\mu^1+\mu^3,\la_2-\la^2+\mu_2-\mu^2\}
  &\leq&\nu^1-\nu^3\nn
 \leq{\rm min}\{-\la^1+\la^3+\mu^2,
  \la^2-\mu^1+\mu^3,\la^2-\la_2+\mu^2-\mu_2\}&.&
\label{eqfour}
\eea

A recent discussion \cite{Zel}
includes a brief review of the problem of determining
when a tensor product multiplicity is non-vanishing. 
The focus there is on $gl(N)$ (and therefore also on $su(N)$) and results
are provided in the form of polyhedral combinatorial expressions.
Our prescription described above is in general
more explicit than results previously obtained. For lower rank $su(N)$,
though, the explicit inequalities may be obtained using various 
alternative approaches.

\subsection{A refinement}

Here we shall indicate how one may derive sets of inequalities determining
when a tensor product multiplicity is greater than a given non-negative
integer $K$
\ben
 T_{\la,\mu,\nu}>K\ .
\label{K}
\een
The case $K=0$ was discussed above.

Our approach is straightforward since the problem translates into
studying when a convex polytope has a (discretised) volume bigger than $K$.
To illustrate, let us
consider $su(3)$. In this case the volume is expressed as a single
sum (\ref{Tthree}), so (\ref{K}) is equivalent to requiring 
\bea
 &&{\rm min}\{\mu_1,\la_2,\la^2+\mu^2-\nu^1,
  \la^1+\mu^1-\nu^2,-\la^1+\la^2+\mu^1-\nu^1+\nu^2,\la^2+\mu^1-\mu^2
  +\nu^1-\nu^2\}\nn
 &-& {\rm max}\{0,\la^2+\mu^1-\mu^2-\nu^1,-\la^1+\la^2+\mu^1-\nu^2\}\nn
 &\geq&K\ .
\label{Kthree}
\eea
This leaves us the following 18 inequalities refining (\ref{eqthree})
\bea
 K&\leq&\la_i,\mu_i,\nu_i\ , \ \ \ \ \ {\rm for} \ \ i=1,2\ ,\nn
  {\rm max}\{\la^1-\la^2+\mu^1-\mu^2,-\la^1+\mu^2,\la^2-\mu^1\}+K&\leq
 &\nu^1\leq\la^2+\mu^2-K\ ,\nn
 {\rm max}\{-\la^1+\la^2-\mu^1+\mu^2,\la^1-\mu^2,-\la^2+\mu^1\}+K&\leq
 &\nu^2\leq\la^1+\mu^1-K\ ,\nn
 {\rm max}\{-\la^2-\mu^1+\mu^2,-\la^1+\la^2-\mu^2\}+K&\leq
 &\nu^1-\nu^2\nn
 &\leq&{\rm min}\{\la^1-\mu^1+\mu^2,\nn
 &&\hspace{1cm}-\la^1+\la^2+\mu^1\}-K\ .
\label{refthree}
\eea
To the best of our knowledege, this is a new result.

In the case $su(4)$ the situation is already much more complicated.
That is because the polytope is three-dimensional, and we cannot immediately
use the triple-sum formula (\ref{Tfour}). We recall that our 
simplifying method for obtaining an appropriate order of summation,
may include redundant summations contributing zero to the
final expression. We might therefore lose crucial information if we
only considered the multiple sum formula. The remedy is to consider
the original polytope, and require the defining faces to 
embrace a volume of at least the desired value.
We would then be led to consider three-dimensional partitions of $K+1$, 
which is beyond the scope of the present work. 
For lower values of $K$ the problem is straightforward, though.

\section{Conclusion}

By virtue of virtual BZ triangles we have obtained a polyhedral combinatorial
expression for the $su(N)$ tensor product multiplicities, different from the
ones discussed in Refs. \cite{GZ,BZmath}.
The main merit of our expression is that it admits a simple measurement
of the convex polytope volume in terms of a multiple sum formula.
The latter is then a new and explicit way of expressing the tensor
product multiplicities of $su(N)$. 

As an application, one may derive explicit bounds on a triple of weights
determining when the associated coupling to the singlet exists. 
To illustrate, the bounds were written for $su(3)$ and $su(4)$. 
Also included was a brief discussion on how to generalise this to bounds
describing $T_{\la,\mu,\nu}>K$.

We believe that our multiple sum representation of the tensor product
multiplicities provides a significant computational improvement over
previous (combinatorial) results. In particular, it is expected to
lead to considerable simplifications when implemented in computer
programs.

It is our hope that our results may find applications to the computation
of fusion rules in conformal field theory with affine Lie group symmetry,
the so-called WZW theories.
Since tensor product multiplicities correspond to the infinite-level
limit of fusion multiplicities, it is helpful to have simple
descriptions of the former in order to understand the latter.
\\[.3cm]
{\it Acknowledgements}\\[.2cm]
We are grateful to G. Flynn for discussions and to C. Cummins and P. Mathieu
for commenting on the manuscript.

\end{document}